\documentclass[12pt]{article}
\pdfoutput=1 
\usepackage{slashed}
\usepackage{color, verbatim}
\usepackage{latexsym}
\usepackage{amssymb}
\usepackage{amsmath}
\usepackage{graphicx}
\graphicspath{{figs/}}
\usepackage{arydshln}
\usepackage[dvipsnames]{xcolor}
\usepackage{adjustbox}
\usepackage{easybmat}
\usepackage{bbm}
\usepackage{cite}
\usepackage{bm}
\usepackage{adjustbox}
\usepackage{booktabs}
\usepackage{multirow}
\usepackage{rotating}
\usepackage{mathtools}
\newcommand{\vtext}[1]{\begin{sideways}\small{#1}\end{sideways}}
\newcommand{\MM}{\mathcal{M}}
\DeclareMathOperator{\tr}{Tr}
\renewcommand{\theequation}{\thesection.\arabic{equation}}
\makeatletter
\@addtoreset{equation}{section}
\g@addto@macro\bfseries{\boldmath}
\newcommand\Label[1]{&\refstepcounter{equation}(\theequation)\ltx@label{#1}&}
\makeatother
\allowdisplaybreaks

\begin{document}
%%%%%%%%%%%%%%%%%%%%%%%%%%%%%%
%
\thispagestyle{empty}
\begin{flushright}
%IPPP/19/XX \\
\end{flushright}
\vspace{0.8cm}

\begin{center}
{\Large\sc One-loop running of dimension-six\\[0.1cm]
Higgs-neutrino operators and\\[0.3cm]
implications of a large neutrino dipole moment}
\vspace{0.8cm}

\textbf{
Mikael Chala$^{\,a}$ and Arsenii Titov$^{\,b}$}\\
\vspace{1.cm}
{\em {$^a$CAFPE and Departamento de F\'isica Te\'orica y del Cosmos,
Universidad de Granada, Campus de Fuentenueva, E--18071 Granada, Spain}}\\[0.2cm]
{\em {$^b$Dipartimento di Fisica e Astronomia ``G. Galilei'', Universit\`a degli Studi di Padova \\
and INFN, Sezione di Padova, Via Francesco Marzolo 8, I--35131 Padova, Italy}}\\[0.2cm]
\vspace{0.5cm}
\end{center}
\begin{abstract}
We compute the one-loop running of the dimension-six CP-even Higgs operators in 
the Standard Model 
effective field theory involving the right-handed component of the would-be 
Dirac neutrinos. Then, on the basis of naturalness arguments, for some 
operators we obtain bounds that surpass direct constraints by orders of 
magnitude.
We 
also discuss the implications of a large Dirac neutrino magnetic dipole moment. 
In particular, we demonstrate that a neutrino magnetic moment explaining the 
recent XENON1T excess induces Higgs and $Z$ invisible decays with branching 
ratios in the range $[10^{-18}, 10^{-12}]$. These numbers are unfortunately 
beyond the reach of current and near future facilities.
\end{abstract}

\newpage

%\tableofcontents

%%%%%%%%%%%%%%%%%%%%%%%%%%%%%%
\section{Introduction}
%%%%%%%%%%%%%%%%%%%%%%%%%%%%%%
%
The Standard Model (SM) effective field theory (EFT)~\cite{Buchmuller:1985jz,Grzadkowski:2010es} is the right tool to describe physics above
the electroweak (EW) scale. Its use has been boosted in the last years~\cite{Brivio:2017vri} in light of
the null results (modulo a few non-conclusive anomalies~\cite{Khachatryan:2015isa,Sirunyan:2017dhj,Aprile:2020tmw}) in the search for new physics at different facilities, and in particular at the LHC. The necessity of using this framework
across a wide range of energies
has also triggered the computation of the one-loop renormalisation group equations (RGEs)
for the dimension-six operators~\cite{Grojean:2013kd,Elias-Miro:2013gya,Elias-Miro:2013mua,Jenkins:2013zja,Jenkins:2013wua,Alonso:2013hga,Liao:2019tep}. The RGEs in the theory valid at energies below the EW scale
where the top quark, the Higgs and the $W$ and $Z$ gauge bosons are integrated out, usually referred to as LEFT, 
are also known~\cite{Jenkins:2017dyc}; as well as the matching between the SMEFT and the LEFT at up to one loop~\cite{Jenkins:2017jig,Dekens:2019ept}.
Likewise, the desire of connecting the SMEFT to ultraviolet (UV)
models has stimulated different works on the matching procedure~\cite{Henning:2014wua,Henning:2016lyp,Ellis:2016enq,Fuentes-Martin:2016uol,Zhang:2016pja,Ellis:2017jns,delAguila:2016zcb,Criado:2017khh,Bakshi:2018ics,Brivio:2019irc}; including more recently the 
first basis of dimension-six operators suitable for off-shell integration~\cite{Gherardi:2020det}.

All the aforementioned works assume that neutrinos are Majorana fermions. Notwithstanding the good motivation
for this option---in particular lepton number (LN) is only an approximate symmetry of the
renormalisable SM Lagrangian---it should not be forgotten that there is absolutely no experimental evidence that
neutrinos are not just Dirac particles as all the other SM fermions. There is even theoretical support for this~\footnote{For example, Refs.~\cite{Ibanez:2017kvh,ArkaniHamed:2007gg} show that the SM with only $3$ Majorana neutrinos does 
satisfy the sharpened version of the weak gravity conjecture by Ooguri and Vafa~\cite{Ooguri:2016pdq}, presumably implying that
 such SM cannot be consistently embedded into a quantum theory of gravity.}.
However, the SMEFT that includes the right-handed (RH) neutrinos $N$, also known as NSMEFT~\cite{delAguila:2008ir,Liao:2016qyd}, has been explored to a smaller extent; see Refs.~\cite{Duarte:2015iba,Duarte:2016caz,Alcaide:2019pnf,Butterworth:2019iff,Cai:2017mow,Bischer:2019ttk,Dekens:2020ttz,Bolton:2020xsm,Duarte:2020vgj} for phenomenological works. The off-shell basis of the NSMEFT
has only recently been worked out in Ref.~\cite{Chala:2020vqp}, where the NLEFT and the tree-level matching between the
two EFTs are also presented. More importantly, only the gauge dependence of the RGEs of only very small set of operators are known~\cite{Bell:2005kz,Han:2020pff}.

Our aim in this paper is to compute the one-loop RGE matrix of the NSMEFT Higgs operators 
in full detail and to discuss the phenomenological implications, particularly in light of the recent XENON1T observation of an excess of low-energy electron recoil events~\cite{Aprile:2020tmw}. 

The article is organised as follows. In section~\ref{sec:nsmeft} we introduce
the NSMEFT and discuss the generic structure of the RGEs. In section~\ref{sec:matching}
we thoroughly discuss the matching of the UV divergences onto the EFT. We obtain
  the corresponding
counterterms and derive our main result, namely the $5\times 5$ anomalous dimension matrix to one loop,
in section~\ref{sec:renormalisation}. In section~\ref{sec:pheno} we discuss some phenomenological implications. In particular, 
the aforementioned XENON1T anomaly might point out to a large neutrino magnetic dipole moment; 
we demonstrate that it leads to irreducible Higgs and $Z$ invisible decays
and we quantify their magnitude.
We conclude in section~\ref{sec:conclusions}, while Appendix~\ref{app:xchecks} is dedicated to different cross-checks 
of our computation.

%%%%%%%%%%%%%%%%%%%%%%%%%%%%%%
\section{The lepton number conserving Standard Model effective field theory}
\label{sec:nsmeft}
%%%%%%%%%%%%%%%%%%%%%%%%%%%%%%
%
We denote by $e, u$ and $d$ the RH leptons and quarks; and by 
$L$ and $Q$ the left-handed (LH)
counterparts. The gluon and the EW gauge bosons are named by $G$ and $W, B$, 
respectively. We represent the Higgs doublet by $H = (H^+, H_0)^T$, 
and $\tilde{H} = i\sigma_2 H^*$,  with $\sigma_I$, $I=1,2,3$, being the Pauli matrices.

Our conventions for the covariant derivative and for
the field strength tensors are
\begin{equation}
 D_\mu = \partial_\mu - ig_1 Y 
B_\mu - i g_2 \frac{\sigma^I}{2} W^I_\mu -i g_s\frac{\lambda^A}{2} G^A_\mu \,,
\end{equation} 
and
\begin{align}
 B_{\mu\nu} &= \partial_\mu B_\nu - \partial_\nu B_\mu\,, \\
W_{\mu\nu}^I &= \partial_\mu W_\nu^I-\partial_\nu W_\mu^I + g_2 \varepsilon^{IJK} W_\mu^J W_\nu^K\,, \\
G_{\mu\nu}^A &= \partial_\mu G_\nu^A-\partial_\nu G_\mu^A + g_s f^{ABC} 
G_\mu^B G_\nu^C\,,
\end{align}
where $Y$ stands for the hypercharge and $\lambda^A$, $A=1,...,8$, are
the Gell-Mann
matrices; while $\epsilon^{IJK}$ and $f^{ABC}$ represent the $SU(2)_L$ and $SU(3)_c$
structure constants.

We denote by $N$ the RH component of the neutrino. 
The renormalisable Lagrangian of the NSMEFT
reads
\begin{align}
 \mathcal{L}_{4} &= -\frac{1}{4}G_{\mu\nu}^A G^{A \mu\nu} -\frac{1}{4} 
W_{\mu\nu}^I W^{I \mu\nu} -\frac{1}{4}B_{\mu\nu}B^{\mu\nu} \nonumber\\
 &\phantom{{}={}}+ \left(D_\mu H\right)^\dagger \left(D^\mu H\right) + \mu_H^2 
H^\dagger H -\frac{1}{2}\lambda_H \left(H^\dagger H\right)^2\nonumber\\
 &\phantom{{}={}}+i\left(\overline{Q}\slashed{D} Q + \overline{u}\slashed{D} u + \overline{d}\slashed{D} d + \overline{L}\slashed{D} L +\overline{e}\slashed{D} e +\overline{N}\slashed{D}N\right) \nonumber\\
 &\phantom{{}={}} - \left[ \overline{Q} Y_d H d + \overline{Q} Y_u \tilde{H}u 
+ \overline{L}Y_e H e + \overline{L} Y_N\tilde{H} N + \mathrm{h.c.}\right].
\end{align}
The dimension-six interactions,
\begin{equation}
 \mathcal{L}_{6} = \frac{1}{\Lambda^2}\sum_i \alpha_i \mathcal{O}_i\,,
\end{equation}
can be expressed in terms of a basis of effective operators. 
We choose the latter to consist of
the SMEFT operators in Ref.~\cite{Grzadkowski:2010es} 
(which do not contain $N$) plus those in Tabs.~\ref{tab:bosonic} 
and~\ref{tab:fermionic}. The $\alpha_i$ represent Wilson coefficients. As we enforce
LN conservation, there are
no dimension-five operators.
\begin{table}[t]
\renewcommand{\arraystretch}{1.5}
 \centering
 \adjustbox{width=\textwidth}{
 \begin{tabular}{|c|c|c|c}\hline
  $0-$Higgs & $1-$Higgs & $2-$Higgs \\
  \hline
  \textcolor{gray}{$\mathcal{O}_{DN}^1=\overline{N} \partial^2 \slashed{\partial} 
N$} & $\mathcal{O}_{NB} = \overline{L}\sigma^{\mu\nu} 
N \tilde{H} 
B_{\mu\nu}$, $\mathcal{O}_{NW}=\overline{L}\sigma^{\mu\nu} 
N\sigma_I\tilde{H} W_{\mu\nu}^I$ & $\mathcal{O}_{HN} = \overline{N}\gamma^\mu 
N (H^\dagger i D_\mu 
H)$  \\[0.2cm]
\textcolor{gray}{$\mathcal{O}_{DN}^2= i \tilde B_{\mu\nu} 
(\overline{N}\gamma^\mu \partial^\nu N)$} & \textcolor{gray}{$\mathcal{O}_{LN}^1=\overline{L} N D^2\tilde{H}$ }, \textcolor{gray}{$\mathcal{O}_{LN}^2=\overline{L} \partial_\mu N D^\mu 
\tilde{H}$ } & \textcolor{gray}{$\mathcal{O}_{NN}^2=\overline{N} i\slashed{\partial} N(H^\dagger 
H)$} \\[0.2cm]
\textcolor{gray}{$\mathcal{O}_{DN}^3=\partial^\nu B_{\mu\nu} (\overline{N}\gamma^\mu 
N)$} & \textcolor{gray}{$\mathcal{O}_{LN}^3=i\overline{L}\sigma^{\mu\nu} \partial_\mu N D_\nu 
\tilde{H}$ }, \textcolor{gray}{$\mathcal{O}_{LN}^4=\overline{L} (\partial^2 N)\tilde{H}$ } & $\mathcal{O}_{HNe} = \overline{N}\gamma^\mu e (\tilde{H}^\dagger iD_\mu H)$\\[0.2cm]
  \hline
  \multicolumn{3}{|c|}{$3-$Higgs: $\mathcal{O}_{LNH} = \overline{L}\tilde{H} N (H^\dagger H)$}\\
  \hline
 \end{tabular}
 }
  \caption{\it Relevant CP-even bosonic operators. 
  The h.c.~is implied when needed. For example, $\mathcal{O}_{DN}^1 = 
\overline{N}\partial^2\slashed{\partial}N +\text{h.c.}$ So all Wilson coefficients are
hermitian. The CP-odd operators include $iB_{\mu\nu} (\overline{N}\gamma^\mu\partial^\nu N)$, 
$i\mathcal{O}_{NB}$, $i\mathcal{O}_{NW}$, $i\mathcal{O}_{LN}^{1,2,3,4}$, $i\mathcal{O}_{LNH}$, $i\mathcal{O}_{HN}$ and $i\mathcal{O}_{HNe}$~\cite{Chala:2020vqp}.
  }\label{tab:bosonic}
\end{table}
\begin{table}[t]
\vspace{1cm}
\renewcommand{\arraystretch}{1.5}
\centering
\begin{tabular}{|c c c|}
\hline
% 4F: RRRR
\multirow{3}{*}{\vtext{RRRR}}&\multicolumn{2}{c|}{$\mathcal{O}_{NN}=(\overline{N}\gamma_\mu N)(\overline{N}\gamma^\mu N)$} \\
&${\cal O}_{eN}=(\overline{e}\gamma_\mu e)(\overline{N}\gamma^\mu N)$&${\cal O}_{uN}=(\overline{u}\gamma_\mu u)(\overline{N}\gamma^\mu N)$\\
&${\cal O}_{dN}=(\overline{d}\gamma_\mu d)(\overline{N}\gamma^\mu N)$&${\cal O}_{duNe}=(\overline{d}\gamma_\mu u)(\overline{N}\gamma^\mu e)$\\
\hline
% 4F: LLRR
LLRR&${\cal O}_{LN}=(\overline{L}\gamma_\mu L)(\overline{N}\gamma^\mu N)$&${\cal O}_{QN}=(\overline{Q}\gamma_\mu Q)(\overline{N}\gamma^\mu N)$\\
\hline
% 4F: LRLR
\multirow{2}{*}{\vtext{LRLR}}&${\cal O}_{LNLe}=(\overline{L} N)\epsilon (\overline{L}e)$&${\cal O}_{LNQd}=(\overline{L} N)\epsilon (\overline{Q} d)$\\
& \multicolumn{2}{c|}{${\cal O}_{LdQN}=(\overline{L}d)\epsilon (\overline{Q} N)$} \\
\hline
% $4F: LRRL
LRRL & \multicolumn{2}{c|}{${\cal O}_{QuNL}=(\overline{Q}u)(\overline{N}L)$} \\
\hline
\end{tabular}
\caption{\it CP-even four-fermion operators. The CP-odd ones carry an extra imaginary unit.}\label{tab:fermionic}
\end{table}

In this work we are only interested in the CP-even sector of the theory. Therefore,
in good approximation we can assume that $Y_u = \text{diag}(y_u, y_c, y_t)$, while
$Y_d = \text{diag}(y_d, y_s, y_b)$ and $Y_e = \text{diag}(y_e, y_\mu, y_\tau)$
without loss of generality.

In good approximation we can also assume
that there is no huge 
fine-tuning between the operators entering into the expression for the neutrino 
mass, $m_\nu\sim Y_N v - \alpha_{LNH}v^3/\Lambda^2$, so in particular 
$Y_N$ can be neglected~\footnote{Even if, as we show below, $\alpha_{LNH}$ is 
generated radiatively and therefore $Y_N\sim g^2 v^2/(16\pi^2\Lambda^2)$ to keep $m_\nu$ small,
$Y_N$ is of order $\lesssim 10^{-4}$ for $\Lambda=1$ TeV, and hence much smaller than
even the muon Yukawa.}. This also implies that lepton flavour is conserved in 
$\mathcal{L}_4$. For simplicity we focus on the regime in
which lepton flavour is also conserved in the $N$ sector of
$\mathcal{L}_6$. As a consequence, the three lepton families factorise (in 
particular they evolve independently under the RGEs). We can therefore ignore 
flavour indices for clarity.

The operators in grey in Tabs.~\ref{tab:bosonic} and \ref{tab:fermionic} are redundant when evaluated on shell; the redundancies
due to algebraic or Fierz identities or ensuing from integration by parts have 
been 
removed. We refer to this basis as \textit{off-shell} or \textit{Green} basis; 
see Ref.~\cite{Gherardi:2020det} for a Green basis of the sector with no $N$. 

The 
relevant equations of motion of $\mathcal{L}_4$ for the fermions read: 
\begin{align}
 i\slashed{D} L &= Y_e H e + Y_N \tilde{H} N\,,\\
 i\slashed{\partial} N &= Y_N^\dagger \tilde{H}^\dagger L\,,\\
 i\slashed{D} e &= Y_e^\dagger H^\dagger L\,,\\
 i\slashed{D} Q &= Y_u \tilde{H} u + Y_d H d\,,\\
 i\slashed{D} u &= Y_u^\dagger \tilde{H}^\dagger Q\,,\\
 i\slashed{D} d &= Y_d^\dagger H^\dagger Q\,;
\end{align}
while for the bosons we have instead:
\begin{align}
 (D^2\tilde{H})^i &= \mu_H^2 \tilde{H}^i - \lambda_H (H^\dagger H)\tilde{H}^i 
 -\epsilon_{ij} \overline{Q^j} Y_d d -\overline{u} Y_u^\dagger Q^i - 
 \epsilon_{ij} \overline{L^j} Y_e e - \overline{N} Y_N L^i\,,\\
 \partial^\nu B_{\nu\mu} &= -\frac{g_1}{2} (i H^\dagger D_\mu H + \text{h.c.}) - g_1 Y^f \overline{f}\gamma_\mu f\,,\\
 D^\nu W_{\nu\mu}^I &= -\frac{g_2}{2} (H^\dagger i D^I_\mu H- iD^I_\mu H^\dagger H + \overline{L}\gamma_\mu\sigma^I L + \overline{Q}\gamma_\mu\sigma^I Q)\,,\\
 D^\nu G_{\nu\mu}^A &= -\frac{g_s}{2} (\overline{Q}\gamma_\mu \lambda^A Q + \overline{u} \gamma_\mu \lambda^A u+\overline{d}\gamma_d \lambda^A d)\,;
\end{align}
where $f$ runs over all fermions. 
As a consequence, the following relations hold on shell for the operators in 
grey in Tab.~\ref{tab:bosonic}:
\begin{align}
 \mathcal{O}_{DN}^1  &= 0 + \cdots\,,\label{eq:odn1} \\
 \mathcal{O}_{DN}^2 &= -\frac{g_1}{2}\mathcal{O}_{HN} + \cdots\,, \\
 \mathcal{O}_{DN}^3  &= - \mathcal{O}_{DN}^2 + \cdots\,,  \\
 \mathcal{O}_{LN}^1  &= \left(\mu_H^2 
\overline{L}\tilde{H}N+\mathrm{h.c.}\right)-\lambda_H \mathcal{O}_{LNH} 
+ 	\cdots \,, \\
 \mathcal{O}_{LN}^2  &= -\frac{1}{2}Y_e \mathcal{O}_{HN} 
-\left(\frac{\mu_H^2}{2}\overline{L}\tilde{H}N +\mathrm{h.c.}\right) + 
 \frac{\lambda_H}{2}\mathcal{O}_{LNH} 
 -\frac{g_1}{8}\mathcal{O}_{NB}+\frac{g_2}{8}\mathcal{O}_{NW} + \cdots\,,\\
 \mathcal{O}_{LN}^3  &= - \mathcal{O}_{LN}^2 + \cdots \,, \\
 \mathcal{O}_{LN}^4  &= 0 + \cdots\,, \\
 \mathcal{O}_{NN}^2  &= 0 + \cdots\,. \label{eq:onn2}
\end{align}
The ellipses represent $Y_N$ suppressed operators
(which might include CP-odd ones) and/or four-fermions, which we ignore~\footnote{
While loops of 
bosonic operators can generate contact interactions, the latter cannot contribute 
back to bosonic operators via equations of motion and they can therefore be 
consistently ignored.}.

Under RG running the Wilson coefficients evolve as
\begin{equation}
 \vec{\beta} \equiv 16\pi^2 \mu\frac{d\vec{\alpha}}{d\mu} = \gamma \vec{\alpha}\,,
\end{equation}
where $\vec{\alpha}$ is a vector that collects the Wilson coefficients
of the EFT basis and $\gamma$ is the so-called \textit{anomalous dimension matrix}.

Given the previous discussion, we can anticipate the global structure of $\gamma$:
\begin{equation}
\gamma =
\begin{pmatrix}
 \boxed{\gamma^{\rm SMEFT}} &  & & &  & & \\
 & \times & \times  & 0 & 0 & 0\\
 & \times & \times & 0 & 0 & 0\\
 & \times & \times  & \times & \mathcal{O}(Y_e) & \mathcal{O}(Y_e) \\
 & \times & \times  & \mathcal{O}(Y_e)  & \times & \mathcal{O}(Y_e) \\
 & \times & \times  & \mathcal{O}(Y_e)  & \mathcal{O}(Y_e) & \times\\
\end{pmatrix}\label{eq:gamma}
\begin{matrix*}[l]
 \phantom{\boxed{\gamma^{\rm SMEFT}}}\\
 \scriptstyle\alpha_{NB}\\
 \scriptstyle\alpha_{NW}\\
 \scriptstyle\alpha_{HN}\\
 \scriptstyle\alpha_{HNe}\\
 \scriptstyle\alpha_{LNH}
\end{matrix*}\,,
\end{equation}
where $\gamma^{\rm SMEFT}$ stands for the $59\times 59$ 
matrix (ignoring flavour indices) accounting for the RG evolution of purely
SMEFT operators~\cite{Jenkins:2013zja,Jenkins:2013wua,Alonso:2013hga}. Because we neglect 
$Y_N$, $N$ does not interact with any other field at the renormalisable level. 
Therefore, operators involving $N$ cannot renormalise purely SMEFT operators, 
and vice versa. This explains the block-diagonal form of the matrix in Eq.~\eqref{eq:gamma}.

The almost diagonal structure in the block of $\lbrace 
\mathcal{O}_{HN},\mathcal{O}_{HNe},\mathcal{O}_{LNH} \rbrace$, only broken by 
$Y_e$, can be explained as follows. Let us define the symmetries 
$\mathbb{L}_e: e\to \exp{(i\theta_e)} e$, $\mathbb{L}_N: N\to 
\exp{(i\theta_N)}N$ and $\mathbb{L}_H: H\to \exp{(i\theta_H)}H$. In the limit 
$Y_e\to 0$, $\mathcal{L}_4$ is completely invariant under the simultaneous 
action of $\mathbb{L}_e$, $\mathbb{L}_N$ and $\mathbb{L}_H$, therefore 
dimension-four loop corrections to dimension-six operators cannot modify the 
$e$, $N$ and Higgs numbers; not even by equations of motion. However,  
$\mathcal{O}_{HN}$, $\mathcal{O}_{HNe}$ and $\mathcal{O}_{LNH}$ differ among 
themselves in at least one of these quantum numbers.

The main result of this paper is the exact one-loop expression for the $5\times 
5$ lower block in $\gamma$, the calculation of which we discuss in detail in
the next 
sections.

%%%%%%%%%%%%%%%%%%%%%%%%%%%%%%
\section{Computation of the divergences}
\label{sec:matching}
%%%%%%%%%%%%%%%%%%%%%%%%%%%%%%
%
We use the background field method. Each of the gauge bosons is thus split
into a background field and a quantum fluctuation that can only appear
in loops in Feynman diagrams.
We work in the Feynman gauge. The
latter is fixed only with respect to the quantum fluctuations, therefore
even non-physical quantities such as counterterms are manifestly
gauge invariant. Consequently, to order $\mathcal{O}(1/\Lambda^2)$,
any one-loop amplitude (and the divergences themselves) can be
unambiguously mapped onto the EFT basis
of Tab.~\ref{tab:bosonic}. Note also that because 
this Green basis contains
operators related by field redefinitions, we can restrict our
calculations to (off-shell) one-particle-irreducible  amplitudes.

Let us first consider the amplitude 
for $\overline{N}(p_1)N (p_2)\to B(p_3)$.
Hereafter we work in dimensional regularisation with space-time
dimension $d=4-2\epsilon$ and absorb $1/\Lambda^2$ in the Wilson coefficients in the expressions for amplitudes. Using \texttt{FeynArts}~\cite{Hahn:2000kx} together
with \texttt{FormCalc}~\cite{Hahn:1998yk} we obtain the following one-loop divergence to order 
$\mathcal{O}(p^2)$: 
\begin{equation}
 i\MM_\mathrm{loop} = \frac{i}{48\pi^2\epsilon} g_1\alpha_{HN}\overline{v_1} 
\left(p_3^2\gamma^\mu -p_3^\mu\slashed{p}_3\right) P_R u_2\epsilon_\mu^\ast\,.\label{eq:div1}
\end{equation}
Here and in what follows $v_1\equiv v(p_1)$, $u_2\equiv u(p_2)$ and 
$\epsilon_\mu^\ast \equiv \epsilon_\mu^\ast(p_3)$.
The most generic divergence in the EFT depends only on $\mathcal{O}_{DN}^2$ and
$\mathcal{O}_{DN}^3$. It reads 
\begin{equation}
 i\mathcal{M}_\mathrm{div} = i \overline{v_1}\left[ \widetilde\alpha_{DN}^3 
\left(p_3^\mu\slashed{p}_3-p_3^2\gamma^\mu\right)
+2\widetilde\alpha_{DN}^2 \left(\gamma^\mu p_2 p_3-\gamma^\mu\slashed{p}_3\slashed{p}_2+ p_3^\mu\slashed{p}_2-p_2^\mu\slashed{p}_3\right)\right] P_R u_2 \epsilon_\mu^\ast\,.
\end{equation}
Upon equating $\mathcal{M}_\mathrm{loop}$ and $\mathcal{M}_\mathrm{div}$, we obtain:
\begin{align}
 \widetilde\alpha_{DN}^2 & = 0\,,\\
 \widetilde\alpha_{DN}^3 &= -\frac{1}{48\pi^2\epsilon}g_1\alpha_{HN}\,.
\end{align}

For the amplitude for $\overline{\nu_L}(p_1) N (p_2)\to H_0(p_3)$ at 
$\mathcal{O}(p^2)$ we obtain:
\begin{equation}
i\MM_\mathrm{loop} = \frac{i}{32\pi^2\epsilon}\overline{v_1}\left[\left(3g_1\alpha_{NB}-9g_2\alpha_{NW} + 2 Y_e \alpha_{HNe}\right) p_1^2 + Y_e\alpha_{HNe}\slashed{p}_1\slashed{p}_2\right] P_R u_2\,,\label{eq:div2}
\end{equation}
and
\begin{align}\nonumber
 i\mathcal{M}_\text{div} = i \overline{v_1} \bigg[\widetilde\alpha_{LN}^1 p_1^2 
 &+\left(\widetilde\alpha_{LN}^1-\widetilde\alpha_{LN}^2+\widetilde\alpha_{LN}^4\right) p_2^2\\
 &+ \left(2\widetilde\alpha_{LN}^1-\widetilde\alpha_{LN}^2+\widetilde\alpha_{LN}^3\right) p_1 p_2  - 
\widetilde\alpha_{LN}^3\slashed{p}_1\slashed{p}_2\bigg] P_R u_2\,,
\end{align}
which implies
\begin{align}
 \widetilde\alpha_{LN}^1 &= \frac{1}{32\pi^2\epsilon}
 \left(3g_1\alpha_{NB}-9g_2\alpha_{NW}+2Y_e\alpha_{HNe}\right), \\
 \widetilde\alpha_{LN}^2 &= \frac{3}{32\pi^2\epsilon}
 \left(2g_1\alpha_{NB}-6g_2\alpha_{NW}+Y_e\alpha_{HNe}\right), \\
 \widetilde\alpha_{LN}^3 &= -\frac{1}{32\pi^2\epsilon}Y_e\alpha_{HNe}\,, \\
 \widetilde\alpha_{LN}^4 &= \frac{1}{32\pi^2\epsilon}
 \left(3g_1\alpha_{NB}-9g_2\alpha_{NW}+Y_e\alpha_{HNe}\right).
\end{align}

For the amplitude for $\overline{\nu_L}(p_1)N(p_2)\to B(p_3) H_0(p_4)$ at 
linear order in the external momenta we get:
\begin{align}\nonumber
 i\MM_\mathrm{loop} &= \frac{i}{64\pi^2\epsilon}\overline{v_1} \bigg\lbrace 2\left(3g_1^2\alpha_{NB}-9g_1g_2\alpha_{NW}+2 g_1 Y_e \alpha_{HNe}\right)p_1^\mu\\\nonumber
 &\phantom{{}={}}+ 2\left[\left(3g_2^2-2g_1^2+4Y_e^2\right)\alpha_{NB}+9g_1g_2\alpha_{NW}\right]p_3^\mu 
 +g_1Y_e\alpha_{HNe}\gamma^\mu\slashed{p}_2\\
 &\phantom{{}={}}+ \left[\left(g_1^2-6g_2^2-8Y_e^2\right)\alpha_{NB} - 9g_1g_2\alpha_{NW}-2g_1Y_e\alpha_{HNe}\right]\gamma^\mu \slashed{p}_3 \bigg\rbrace P_R u_2\epsilon_\mu^\ast\,,
 \label{eq:div3}
\end{align}
as well as
\begin{align}\nonumber
 i\MM_\mathrm{div} = ig_1\overline{v_1} \bigg[\widetilde\alpha_{LN}^1 p_1^\mu &+ \left(\widetilde\alpha_{LN}^1-\frac{1}{2}\widetilde\alpha_{LN}^2+\frac{1}{2}\widetilde\alpha_{LN}^3\right) p_2^\mu
 + \left(2\frac{\widetilde\alpha_{NB}}{g_1}-\frac{1}{2}\widetilde\alpha_{LN}^1\right) p_3^\mu \\
 &-\frac{1}{2}\widetilde\alpha_{LN}^3\gamma^\mu\slashed{p}_2 - 2\frac{\widetilde\alpha_{NB}}{g_1}\gamma^\mu\slashed{p}_3\bigg] P_R u_2 \epsilon_\mu^\ast\,.
\end{align}
Upon equating both quantities, we obtain the same values of $\widetilde\alpha_{LN}^1$, $\widetilde\alpha_{LN}^2$ and $\widetilde\alpha_{LN}^3$ as before (what provides a strong cross-check of the computation), as well as
\begin{equation}
 \widetilde\alpha_{NB} = \frac{1}{128\pi^2\epsilon}
 \left[\left(6g_2^2-g_1^2+8Y_e^2\right)\alpha_{NB} +9g_1g_2\alpha_{NW}
 +2g_1Y_e\alpha_{HNe}\right].
\end{equation}

The divergences at one loop and in the EFT for $\overline{e_L}(p_1) 
N(p_2)\to W^3 (p_3) H^+ (p_4)$ at $\mathcal{O}(p)$ read respectively:
\begin{align}\nonumber
 i\MM_\mathrm{loop} = \frac{i}{64\pi^2\epsilon}\overline{v_1}\bigg\lbrace& 2g_2\left(9g_2\alpha_{NW}-3g_1\alpha_{NB}-2  Y_e\alpha_{HNe}\right)p_1^\mu\\\nonumber
 &+ 2\left[3g_1g_2\alpha_{NB}+\left(g_1^2-6g_2^2-4 Y_e^2\right)\alpha_{NW}\right]p_3^\mu\\\nonumber
 &-g_2Y_e\alpha_{HNe}\gamma^\mu\slashed{p}_2\\
 &-\left[3g_1g_2\alpha_{NB} 
 + \left(2g_1^2-3g_2^2-8 Y_e^2\right)\alpha_{NW}
 - 2 g_2 Y_e \alpha_{HNe}\right] \gamma^\mu \slashed{p}_3\bigg\rbrace P_R u_2\epsilon_\mu^\ast\,,
 \label{eq:div4}
\end{align}
and
\begin{align}\nonumber
 i\MM_\mathrm{div} = ig_2\overline{v_1} \bigg[-\widetilde\alpha_{LN}^1 p_1^\mu & -\left(\widetilde\alpha_{LN}^1-\frac{1}{2}\widetilde\alpha_{LN}^2+\frac{1}{2}\widetilde\alpha_{LN}^3\right)  p_2^\mu
+ \left(2\frac{\widetilde\alpha_{NW}}{g_2}+\frac{1}{2}\widetilde\alpha_{LN}^1\right) p_3^\mu \\
 &+\frac{1}{2}\widetilde\alpha_{LN}^3\gamma^\mu\slashed{p}_2 - 2\frac{\widetilde\alpha_{NW}}{g_2}\gamma^\mu\slashed{p}_3\bigg] P_R u_2 \epsilon_\mu^\ast\,.
\end{align}
This cross-checks again $\widetilde\alpha_{LN}^1$, $\widetilde\alpha_{LN}^2$ and $\widetilde\alpha_{LN}^3$ and also leads to
\begin{equation}
 \widetilde\alpha_{NW} = \frac{1}{128\pi^2\epsilon}
 \left[3g_1g_2\alpha_{NB} +\left(2g_1^2-3g_2^2-8Y_e^2\right)\alpha_{NW}
 -2g_2Y_e\alpha_{HNe}\right].
\end{equation}

For $\overline{N}(p_1) N(p_2)\to H_0^*(p_3) H_0(p_4)$ to $\mathcal{O}(p)$ 
in the external momenta, we have:
\begin{align}
 i\MM_\mathrm{loop} = \frac{i}{32\pi^2\epsilon}
 \left(g_1^2 + 3g_2^2\right) \alpha_{HN}  
 \overline{v_1}\left(\slashed{p}_1+\slashed{p}_2-2\slashed{p}_3\right)u_2\,,\label{eq:div5}
\end{align}
and
\begin{align}
 i\MM_\mathrm{div} &= i \overline{v_1}\bigg[\left(\widetilde\alpha_{NN}^2-\widetilde\alpha_{HN}\right)\slashed{p}_1 
-\left(\widetilde\alpha_{NN}^2+\widetilde\alpha_{HN}\right)\slashed{p}_2 + 2\widetilde\alpha_{HN}\slashed{p}_3 \bigg]P_R 
u_2\,.
\end{align}
From equating these two amplitudes, we obtain the conditions:
\begin{align}
 \widetilde\alpha_{NN}^2 &= 0\,,\\
 \widetilde\alpha_{HN} &= -\frac{1}{32\pi^2\epsilon}\left(g_1^2+3 g_2^2\right)\alpha_{HN}\,.
\end{align}

The one-loop divergence for $\overline{\nu_L}(p_1)N(p_2) 
H_0^*(p_3)\to H_0^*(p_4) H_0(p_5)$ at zero momentum reads
\begin{align}\nonumber
 i\MM_\mathrm{loop} = \frac{i}{16\pi^2\epsilon}\overline{v_1}\bigg[&
 \left(12\lambda_H-g_1^2-3g_2^2-2Y_e^2\right)\alpha_{LNH} 
 - 3g_1(g_1^2+g_2^2)\alpha_{NB}\\
 &+ 
3g_2\left(g_1^2+3g_2^2+4Y_e^2\right)\alpha_{NW}+Y_e\left(3g_2^2-2\lambda_H-2Y_e^2\right)\alpha_{HNe}\bigg]
P_R u_2\,,\label{eq:div6}
\end{align}
while in the EFT at tree level we have
\begin{align}
 i\MM_\mathrm{div} = -2i\widetilde\alpha_{LNH} \overline{v_1}P_R u_2\,.
 \label{eq:div6b}
\end{align}
This fixes
\begin{align}\nonumber
 \widetilde\alpha_{LNH} = \frac{1}{32\pi^2\epsilon}
 \bigg[&3g_1\left(g_1^2+g_2^2\right)\alpha_{NB}
 -3g_2\left(g_1^2+3g_2^2+4Y_e^2\right)\alpha_{NW} \nonumber \\
 &+Y_e\left(2\lambda_H-3g_2^2+2Y_e^2\right)\alpha_{HNe}
 +\left(g_1^2+3g_2^2-12\lambda_H+2Y_e^2\right)\alpha_{LNH}\bigg]\,.
\end{align}

Finally, upon computing the divergent part of $\overline{N}(p_1)e_R(p_2)\to 
H^-(p_3) H_0^*(p_4)$ at order $\mathcal{O}(p)$, we obtain:
\begin{align}
 i\MM_\mathrm{loop} = 
 \frac{3i}{32\pi^2\epsilon}
 \left[g_1Y_e\alpha_{NB}-3g_2Y_e\alpha_{NW}+\left(g_1^2-g_2^2\right)\alpha_{HNe}\right]
 \overline{v_1}\left(\slashed{p}_3-\slashed{p}_4\right)P_R u_2\,,
 \label{eq:div7}
\end{align}
and
\begin{align}
 i\MM_\mathrm{div} = i\widetilde\alpha_{HNe} \overline{v_1}
 \left(\slashed{p}_3-\slashed{p}_4\right)P_R u_2\,,
\end{align}
which leads to
\begin{equation}
 \widetilde\alpha_{HNe} = \frac{3}{32\pi^2\epsilon} 
 \left[g_1Y_e\alpha_{NB} - 3g_2Y_e\alpha_{NW}
 +\left(g_1^2-g_2^2\right)\alpha_{HNe}\right].
\end{equation}
We provide a completely independent cross-check of these results in Appendix~\ref{app:xchecks}.

To conclude, for the $Z$ factors of the fields we have:
\begin{align}
 Z_H &= 1 + \frac{1}{32 \pi^2 \epsilon} \left[g_1^2 + 3 g_2^2 - 6\tr\left(Y_u^2 + Y_d^2\right)-2 \tr\left(Y_e^2\right)\right],\\
 Z_L &= 1-\frac{1}{64\pi^2\epsilon} \left(g_1^2 + 3 g_2^2 + 2 Y_e^2\right)\,,\\
 Z_e &= 1-\frac{1}{16\pi^2\epsilon} \left(g_1^2 + Y_e^2\right),\\
 Z_B &= 1 - \frac{41 g_1^2}{96 \pi^2\epsilon}\,,\\
 Z_W &= 1 + \frac{19 g_2^2}{96\pi^2\epsilon}\,.
\end{align}
Note that we use $Y_e$ to refer both to a particular entry of the Yukawa matrix and to this
matrix itself (when it comes inside the trace).

%%%%%%%%%%%%%%%%%%%%%%%%%%%%%%
\section{Anomalous dimensions}
\label{sec:renormalisation}
%%%%%%%%%%%%%%%%%%%%%%%%%%%%%%
%
We remove the redundant operators (those in grey in Tab.~\ref{tab:bosonic}) using the relations in Eqs.~\eqref{eq:odn1}--\eqref{eq:onn2}. This shifts the Wilson coefficients $\alpha_{NB}$, $\alpha_{NW}$ and $\alpha_{HN}$:
\begin{align}
 \widetilde\alpha_{NB} &\rightarrow
 \widetilde\alpha_{NB} - \frac{g_1}{8}\left(\widetilde\alpha_{LN}^2-\widetilde\alpha_{LN}^3\right) 
 = \frac{1}{64\pi^2\epsilon}\left[\left(3g_2^2-2g_1^2+4Y_e^2\right)\alpha_{NB}
 + 9g_1g_2\alpha_{NW}\right], \\
 \widetilde\alpha_{NW} &\rightarrow
 \widetilde\alpha_{NW} + \frac{g_2}{8}\left(\widetilde\alpha_{LN}^2-\widetilde\alpha_{LN}^3\right) 
 = \frac{1}{64\pi^2\epsilon}\left[3g_1g_2\alpha_{NB}
 + \left(g_1^2-6g_2^2-4Y_e^2\right)\alpha_{NW}\right], \\
 \widetilde\alpha_{HN} &\rightarrow
 \widetilde\alpha_{HN} - \frac{g_1}{2}\left(\widetilde\alpha_{DN}^2-\widetilde\alpha_{DN}^3\right) - \frac{Y_e}{2}\left(\widetilde\alpha_{LN}^2-\widetilde\alpha_{LN}^3\right) \nonumber \\
 &\phantom{{}\rightarrow{}}= - \frac{1}{96\pi^2\epsilon}\left[
 9g_1Y_e\alpha_{NB} -27g_2Y_e\alpha_{NW} 
 + \left(4g_1^2+9g_2^2\right) \alpha_{HN} + 6Y_e^2\alpha_{HNe}\right];
\end{align}
$\alpha_{HNe}$ and $\alpha_{LNH}$ (accidentally) remain unchanged.

This way, we fully determine the divergent Lagrangian
\begin{equation}
 \mathcal{L}_\mathrm{div} = \frac{1}{32\pi^2\Lambda^2\epsilon}\vec{\mathcal{O}}^T\cdot \mathcal{C} \cdot\vec{\alpha}\,,
\end{equation}
where the vector $\vec{\mathcal{O}}$ encodes the relevant operators, 
and the matrix $\mathcal{C}$ contains only SM couplings.
We use the latter to fix the counterterms in the NSMEFT Lagrangian
\begin{align}
 \mathcal{L}_6 &= \frac{1}{\Lambda^2}\vec{\alpha}^T \cdot \vec{\mathcal{O}} + \frac{1}{\Lambda^2}\vec{\mathcal{O}^T}\cdot \left(Z_F Z-\mathbbm{1}\right)\cdot \vec{\alpha}\\
 &= \frac{1}{\Lambda^2}\vec{\alpha}^T \cdot \vec{\mathcal{O}} + \frac{1}{32\pi^2\Lambda^2\epsilon}\vec{\mathcal{O}^T}\cdot \left(K_F + K\right)\cdot \vec{\alpha}\,,
\end{align}
where $Z_F$ contains the wave-function renormalisation factors~\footnote{Explicitly, $$Z_F = \text{diag}\left(\sqrt{Z_L Z_H Z_B}, \sqrt{Z_L Z_H Z_W}, Z_H,\sqrt{Z_E}Z_H,\sqrt{Z_L} (Z_H)^{3/2}\right)\,.$$}, and we have introduced $Z = \mathbbm{1} + K/(32\pi^2\epsilon)$ and $Z_F = \mathbbm{1} + K_F/(32\pi^2\epsilon)$. We obtain
\begin{equation}
 K = -\left(\mathcal{C} +K_F\right).
\end{equation}
Following \textit{e.g.} Ref.~\cite{Buchalla:2019wsc}, it can be seen that the anomalous dimension matrix $\gamma$ is simply given by $K$. Thus, we finally get
\begin{equation}
 \hspace{-1.5cm}
 \scriptsize
 \renewcommand{\arraystretch}{1.5}
 \gamma = \begin{pmatrix}  
 \frac{91}{12}g_1^2-\frac{9}{4}g_2^2-\frac{3}{2}Y_e^2+\tr^2& -\frac{9}{2}g_1g_2 & 0 & 0 & 0 \\ 
 -\frac{3}{2}g_1g_2 & -\frac{3}{4}g_1^2-\frac{11}{12}g_2^2+\frac{5}{2}Y_e^2+\tr^2 & 0 & 0 & 0 \\ 
 3g_1Y_e & -9g_2Y_e & \frac{1}{3}g_1^2+2\tr^2 & 2Y_e^2 & 0 \\ 
 -3g_1Y_e & 9g_2Y_e & 0 & -3g_1^2+Y_e^2+2\tr^2 & 0 \\ 
 -3g_1\left(g_1^2+g_2^2\right) & 3g_2\left(g_1^2+3g_2^2+4Y_e^2\right) & 0 & Y_e\left(3g_2^2-2\lambda_H-2Y_e^2\right) & -\frac{9}{4}g_1^2-\frac{27}{4}g_2^2+12\lambda_H-\frac{3}{2}Y_e^2+ 3\tr^2
 \end{pmatrix}
 \begin{matrix*}[l]
  \scriptstyle\alpha_{NB}\\
  \scriptstyle\alpha_{NW}\\
  \scriptstyle\alpha_{HN}\\
  \scriptstyle\alpha_{HNe}\\
  \scriptstyle\alpha_{LNH}
 \end{matrix*}\,,
 \label{eq:gamma2}
\end{equation}
where we have defined 
\begin{equation}
\tr^2 \equiv 3\tr\left(Y_u^2+Y_d^2\right) + \tr\left(Y_e^2\right).
\end{equation}

We notice that, as anticipated in section~\ref{sec:nsmeft}, the operators $\mathcal{O}_{HN}$,
 $\mathcal{O}_{HNe}$ and $\mathcal{O}_{LNH}$ do not mix in the limit of vanishing $Y_e$. Also,
 these operators, which can be generated at tree level in UV completions of the SM, do not
 renormalise $\mathcal{O}_{NW}$ and $\mathcal{O}_{NB}$, which can only arise at one loop.
We also stress that some entries, \textit{e.g.} the renormalisation 
of $\mathcal{O}_{HN}$ by $\mathcal{O}_{NW}$, are enhanced 
with respect to the naive dimensional analysis estimation 
by up to an order of magnitude.

%%%%%%%%%%%%%%%%%%%%%%%%%%%%%%
\section{Some phenomenological implications}
\label{sec:pheno}
%%%%%%%%%%%%%%%%%%%%%%%%%%%%%%
%
Among the variety of phenomenological implications, we would like to explore the possibility
that the excess of low-energy electron recoil events recently observed by XENON1T~\cite{Aprile:2020tmw},
which has triggered a lot of attention~\cite{Takahashi:2020bpq,Kannike:2020agf,Alonso-Alvarez:2020cdv,Fornal:2020npv,Boehm:2020ltd,Harigaya:2020ckz,Bally:2020yid,Su:2020zny,Du:2020ybt,DiLuzio:2020jjp,Dey:2020sai,Bell:2020bes,Chen:2020gcl,AristizabalSierra:2020edu,Buch:2020mrg,Choi:2020udy,Paz:2020pbc,Lee:2020wmh,Primulando:2020rdk,Nakayama:2020ikz,1802726,1802727,1802729,1802687}, is due to a relatively large neutrino magnetic dipole moment.  Following Ref.~\cite{Khan:2020vaf} (see also Ref.~\cite{Boehm:2020ltd}), one can take $\mu_\nu\sim 2\times 10^{-11}\mu_B$, where $\mu_B$
stands for the Bohr magneton. (This explanation necessarily assumes that the strong astrophysical
bounds~\cite{Giunti:2014ixa}, which are subject to a number of uncontrollable uncertainties, can not be taken
at face value.)

The neutrino magnetic moment can also be expressed as~\cite{Bell:2005kz}
\begin{equation}
 \bigg|\frac{\mu_\nu}{\mu_B}\bigg| = \frac{4\sqrt{2}}{\mathrm{e}} \frac{m_e 
v}{\Lambda^2} \alpha_{NA}(v)\,,
\end{equation}
where $\alpha_{NA} = c_W \alpha_{NB}+s_W \alpha_{NW}$, $m_e$ represents the
electron mass and $\mathrm{e}=\sqrt{4\pi\alpha_\mathrm{QED}}$ with $\alpha_\mathrm{QED}$ the
electromagnetic fine-structure constant. The non detection of
new particles at the LHC most likely implies that $\alpha_{NA}$ is generated at
a scale $\Lambda \gtrsim $ TeV. 
In what follows we assume two benchmark values of 
$\Lambda = 1$~TeV and $100$~TeV.
We obtain
\begin{equation}
 \alpha_{NA} \sim 9\times 10^{-6}~\left(9\times 10^{-2}\right) 
 \quad \text{for} \quad \Lambda =1~\text{TeV}~(100~\text{TeV})\,.
\end{equation}
Even if this is the only non-vanishing Wilson coefficient at the high scale,
running down from $\Lambda =1$~TeV~(100~TeV) to the EW scale,
we obtain:
\begin{align}
 |\alpha_{LNH}(v)| &\sim 6\times10^{-8}~\left(2\times10^{-3}\right),\\
 |\alpha_{HN}(v)| &\sim 10^{-9}~\left(6\times10^{-5}\right)\,,\\
 |\alpha_{NZ}(v)|  &\sim 2\times10^{-8}~\left(7\times10^{-4}\right)\,,
\end{align}
where $\alpha_{NZ} = c_W \alpha_{NW}-s_W \alpha_{NB}$. (We have assumed that
all neutrinos have similar magnetic moment, so $\alpha_{HN}$ is only suppressed
by the tau Yukawa.) 

An immediate consequence of this result is that the neutrino masses get a
radiative contribution of order % 
$\delta m_\nu = |\alpha_{LNH}| v^3/(2\sqrt{2}\Lambda^2) 
\sim 3\times10^{2}$~eV ($10^{3}$~eV) for the new physics scale $\Lambda =1$~TeV~(100~TeV). 
Most of this correction must be cancelled by the bare $Y_N$, 
implying a fine-tuning of order $\mathcal{O}(10^3-10^4)$.
This observation was already made in Ref.~\cite{Bell:2005kz}. 
The authors of this article obtain the RGEs 
of $\alpha_{NB}$, $\alpha_{NW}$ and $\alpha_{LNH}$
neglecting the Yukawa terms. 
The equivalent block in our $\gamma$ matches their
result up to a factor of $2$ in the mixing of $\alpha_{NB}$ and $\alpha_{NW}$
into $\alpha_{LNH}$, which (slightly) weakens the amount of fine-tuning. 
 Unfortunately, we do not find enough details about the computation in
Ref.~\cite{Bell:2005kz} to disentangle the root of this discrepancy.

Irrespectively of this tuning, given the aforementioned numbers for the Wilson
coefficients and taking into account the three lepton families,
we predict the following Higgs and $Z$ decays for $\Lambda = 1$~TeV (100~TeV):
\begin{equation}
 \Gamma(h\to \text{inv}) = \frac{3 m_h v^4}{16\pi\Lambda^4}\alpha_{LNH}^2
 \sim 9\times10^{-14}~\text{MeV}~\left(2\times10^{-12}~\text{MeV}\right),
\end{equation}
\begin{equation}
 \Gamma(Z\to\text{inv}) = \frac{m_Z^3 v^2}{8 \pi\Lambda^4}(\alpha_{HN}^2+2\alpha_{NZ}^2)
 \sim 10^{-18}~\text{GeV}~\left(2\times10^{-17}~\text{GeV}\right).
\end{equation}
The expected Higgs and $Z$ branching ratios are therefore
\begin{align}
 \mathcal{B}(h\to\text{inv})\sim 2\times 10^{-14}~\left(4\times10^{-13}\right)\,,\\
 \mathcal{B}(Z\to\text{inv})\sim 5\times 10^{-19}~\left(8\times 10^{-18}\right)\,,
\end{align}
where we have used $\Gamma_h^\mathrm{total}\approx 4$~MeV 
and $\Gamma_Z^\mathrm{total}\approx 2.49$~GeV~\cite{Tanabashi:2018oca}.
Unfortunately, these numbers are so small that it is not feasible that they will be tested at any current or future facilities~\cite{Ishikawa:2019uda}.

On a different front, 
from Eq.~\eqref{eq:gamma2} it is also clear that $\mathcal{O}_{HNe}$
generates a contribution $\delta m_\nu$ to the neutrino mass too. For the tau 
flavour,
requiring $\delta m_\nu < 1$ eV, it can be shown that 
$\alpha_{HNe}\lesssim 2\times10^{-6}$ for $\Lambda = 1$~TeV. 
If $\alpha_{HNe}$ is rather generated at
$\Lambda=100$~TeV, we obtain $\alpha_{HNe}\lesssim 2\times10^{-2}$.
These bounds surpass by orders of magnitude the best bound that can
be set on $\alpha_{HNe}$ using measurements of $W$ branching ratios, which
is $\mathcal{O}(1)$~\footnote{Note that assuming $\alpha_{NW}=0$, $$\Delta \Gamma(W\to\ell\nu) = \frac{m_W^3v^2}{48\pi\Lambda^4}\alpha_{HNe}^2\,,$$ while experimentally
this quantity is bounded to $\Delta\Gamma(W\to\ell\nu)/\Gamma_W^\text{total}<2\times 10^{-3}$
at the 95\% CL~\cite{Tanabashi:2018oca}, 
with $\Gamma_W^\text{total}\sim 2.09$ GeV.
Altogether this implies $|\alpha_{HNe}/\Lambda^2|\lesssim 4.5$~TeV$^{-2}$.}.

%%%%%%%%%%%%%%%%%%%%%%%%%%%%%%
\section{Conclusions}
\label{sec:conclusions}
%%%%%%%%%%%%%%%%%%%%%%%%%%%%%%
%
We have computed the RGEs of all dimension-six Higgs operators in the NSMEFT at one loop, 
thereby extending previous partial computations which did not include all the 
operators nor the Yukawa dependence.  Thus, this work comprises a substantial 
step forward towards the description of new physics in terms of EFTs in the 
regime in which neutrinos are Dirac particles.

In our basis, the only operators that do 
not mix among themselves under running are $\mathcal{O}_{HN}$ and $\mathcal{O}_{LNH}$,
while the three operators $\mathcal{O}_{HN}$, $\mathcal{O}_{HNe}$
and $\mathcal{O}_{LNH}$ renormalise independently in the limit 
of vanishing Yukawas (even at higher orders). 

The operators $\mathcal{O}_{NB}$ and $\mathcal{O}_{NW}$, which together
contribute to the neutrino magnetic dipole moment, renormalise 
$\mathcal{O}_{LNH}$ via gauge interactions; all the others are Yukawa 
suppressed. With this in mind, we have also analysed the consequences of 
the recent XENON1T excess~\cite{Aprile:2020tmw} being due to an anomalous Dirac neutrino 
magnetic dipole moment $\mu_\nu\sim 2\times 10^{-11}\mu_B$. We observe that:
\begin{enumerate}
\item A contribution to the neutrino mass of 
order $10^2$--$10^3$~eV would be generated, requiring a sensible cancellation between 
this and the bare mass to account for the tiny observed $m_\nu\sim 0.1$ eV. 
This was already pointed out in Ref.~\cite{Bell:2005kz}. We however
find a small discrepancy with the result in this reference; see section~\ref{sec:pheno}.
\item Irrespectively of whether neutrino masses are tuned, 
$\mathcal{O}_{LNH}$ would be induced radiatively triggering the Higgs 
decay to invisible $h\to \text{inv}$ with branching ratio of order $\sim 
2\times10^{-14}$ ($4\times10^{-13}$) for $\Lambda=1$ TeV ($100$ TeV).
\item If the dipole moment of the electron and tau neutrinos are equally large,
then one also expects a new contribution to the invisible $Z$ decay  with branching
ratio $5\times10^{-19}$ ($8\times10^{-18}$) for $\Lambda=1$ TeV ($100$ TeV).
\end{enumerate}
Unfortunately, even if the XENON1T excess survives in the long
term, these numbers are too small to be explored at
current and near future facilities.

On a different note, we have shown that $\mathcal{O}_{HNe}$ also renormalises the neutrino
mass term by $\delta m_\nu$. Despite being Yukawa suppressed, we find that requiring $\delta m_\nu<1$
eV sets a bound on $\alpha_{HNe}$ orders of magnitude stronger than the current bound based on
limits from $W\to\ell\nu$.

%%%%%%%%%%%%%%%%%%%%%%%%%%%%%%
\section*{Acknowledgements}
%%%%%%%%%%%%%%%%%%%%%%%%%%%%%%
%
We would like to thank Claudius Krause and Jose Santiago 
for useful discussions. 
MC is supported by the Spanish MINECO under the Juan de la Cierva programme as 
well as by the Ministry of Science
and Innovation under grant number FPA2016-78220-C3-3-P, 
and by the Junta de Andaluc{\'\i}a grants FQM 101 and A-FQM-211-UGR18 (fondos FEDER).

%%%%%%%%%%%%%%%%%%%%%%%%%%%%%%
\appendix
\section{Cross-checks}
\label{app:xchecks}
%%%%%%%%%%%%%%%%%%%%%%%%%%%%%%
% 
Our partial yet thorough cross-check consists in computing the gauge 
dependence of UV divergences in the sector of $\alpha_{HN}$, $\alpha_{HNe}$ and $\alpha_{LNH}$, 
evaluating by hand (with the help of \texttt{FeynRules}~\cite{Alloul:2013bka}) each of the diagrams generated with \texttt{QGRAF}~\cite{Nogueira:1991ex}.

We will make use of the following identities~\cite{Chala:2020vqp}: 
\begin{align}
 \int \frac{\mathrm{d}^d k}{(2\pi)^d} \frac{1}{(k^2 - M^2)^n} &= A_n\,,\\\nonumber\\
\int \frac{\mathrm{d}^d k}{(2\pi)^d} \frac{k^\mu k^\nu}{(k^2 - M^2)^n} &= 
g^{\mu\nu}B_n \,,\\\nonumber\\
\int \frac{\mathrm{d}^d k}{(2\pi)^d} \frac{k^\mu k^\nu k^\rho k^\sigma}{(k^2 - M^2)^n} 
&= \left(g^{\mu\nu}g^{\rho\sigma} + g^{\mu\rho} g^{\nu\sigma} + 
g^{\mu\sigma}g^{\nu\rho}\right)C_n\,,
\end{align}
that lead to
\begin{align}
 A_2 &= \frac{i}{16\pi^2\epsilon}+\cdots\,,\\
 B_3 &= \frac{i}{64\pi^2\epsilon}+\cdots\,,\\
 C_4 &= \frac{i}{384\pi^2\epsilon}+\cdots\,,
\end{align}
where the ellipses encode finite terms.

%%%%%%%%%%%%%%%%%%%%%%%%%%%%%%
\subsection{$\overline{N}N\to B$}
%%%%%%%%%%%%%%%%%%%%%%%%%%%%%%
%
The relevant diagrams are given in Fig.~\ref{fig:amp1}. The different contributions to the amplitude read:
\begin{figure}[t]
 \centering
  \includegraphics[height=3cm]{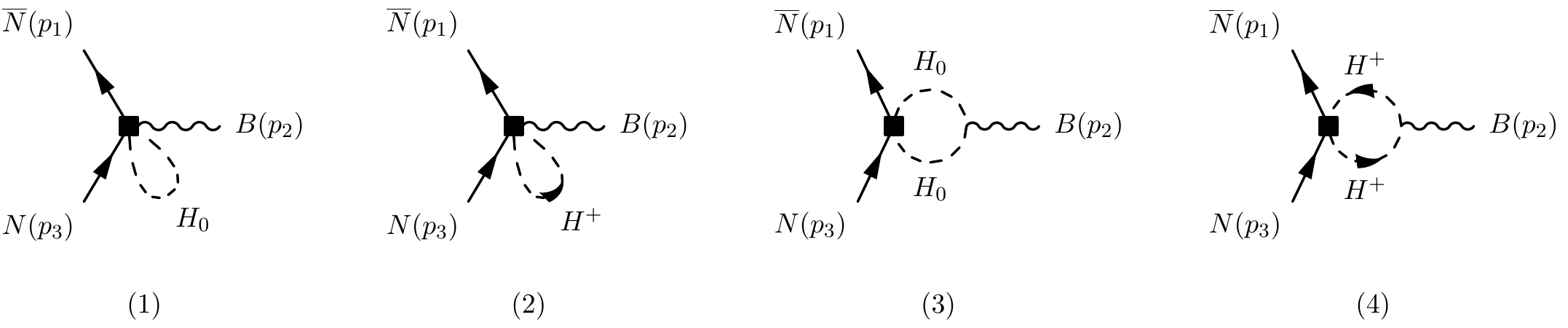}
 \caption{\it Relevant Feynman diagrams for $\overline{N}N\to B$.} 
 \label{fig:amp1}
\end{figure}
\begin{align}
 i\MM_1 &= g_1 \alpha_{HN} A_1 \overline{v_1}\gamma^\mu P_R u_3 \epsilon_\mu^\ast \,,\\
 \MM_2 &= \MM_1 \,,\\
 i\MM_3 &= -\frac{g_1}{2}\alpha_{HN}\overline{v_1}
 \left[4B_2\gamma^\mu + \left(16 C_4-4 B_3\right) p_2^2\gamma^\mu 
 + \left(A_2-8 B_3+32 C_4\right)p_2^\mu\slashed{p}_2\right]P_R u_3\epsilon_\mu^\ast\,,\\
 \mathcal{M}_4 &= \mathcal{M}_3\,.
\end{align}
Here and in what follows $v_1\equiv v(p_1)$, 
$u_3 \equiv u(p_3)$ and $\epsilon_\mu^\ast \equiv \epsilon_\mu^\ast(-p_2)$.

Summing over the four diagrams we obtain
\begin{align}
 i\mathcal{M}_\mathrm{loop} &= \frac{i}{48\pi^2\epsilon} g_1 \alpha_{HN} \overline{v_1} \left(p_2^2 \gamma^\mu - p_2^\mu \slashed{p}_2\right) 
 P_R u_3 \epsilon_\mu^\ast\,.
\end{align}
The \texttt{FeynArts}/\texttt{FormCalc} convention for momenta differs from our setup involving \texttt{QGRAF} and \texttt{FeynRules}. In the former, the momenta of incoming particles
point in and outgoing particles point out, while in the second all momenta point in. The momentum
associated to each particle also differs.

Thus, we have 
 $p_3^\mathrm{FC} = -p_2^\mathrm{QG}$, 
$p_1^\mathrm{FC} = p_1^\mathrm{QG}$ and 
$p_2^\mathrm{FC} = p_3^\mathrm{QG}$. Having this in mind, it is evident that this result agrees
with Eq.~\eqref{eq:div1} in the limit $Y_e, \alpha_{NB},\alpha_{NW}\to 0$.

%%%%%%%%%%%%%%%%%%%%%%%%%%%%%%
\subsection{$\overline{\nu_L} N\to H_0$}
%%%%%%%%%%%%%%%%%%%%%%%%%%%%%%
%
The relevant diagrams are shown in Fig.~\ref{fig:amp2}~\footnote{In what follows the ``missing'' diagrams are either of order $\mathcal{O}(\alpha_{NB,NW})$ or $\mathcal{O}(1/\Lambda^4)$. This is why we do not display them.}. We have:
\begin{figure}[t]
 \centering
  \includegraphics[height=3cm]{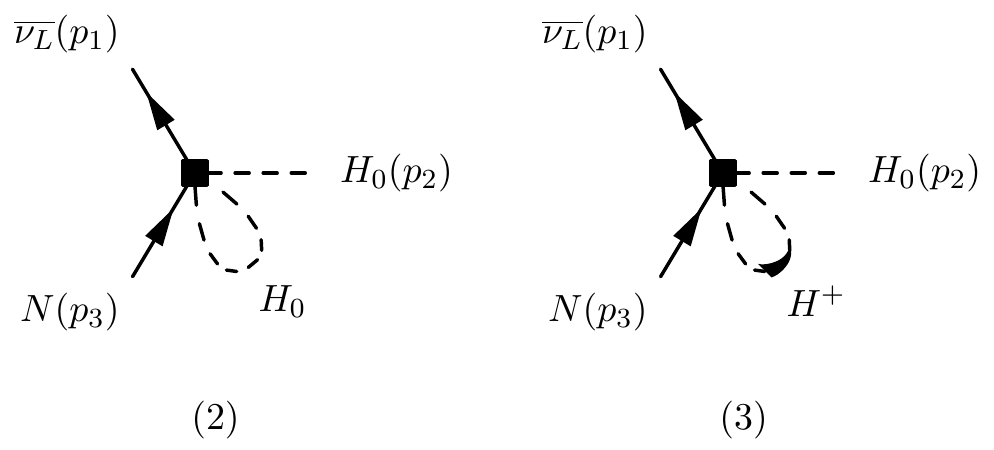}
 \caption{\it Relevant Feynman diagrams for $\overline{\nu_L} N\to H_0$.}
 \label{fig:amp2}
\end{figure}
\begin{align}
 %
 %\mathcal{M}_1 &= \mathcal{O}(\alpha_{NB,NW})\,,\\
 %
 \mathcal{M}_2 &= 0\,,\\
 \mathcal{M}_3 &= 0\,.
 %
 %\mathcal{M}_4 &= \mathcal{O}(\alpha_{NB,NW})\,,\\
 %
 %\mathcal{M}_5 &=\mathcal{O}(\alpha_{NB,NW})\,,\\
 %
 %\mathcal{M}_6 &=\mathcal{O}(\alpha_{NB,NW})\,,\\
 %
 %\mathcal{M}_7 &=\mathcal{O}(\alpha_{NB,NW})\,,\\
 %
 %\mathcal{M}_8 &=\mathcal{O}(\alpha_{NB,NW})\,,\\
 %
 %\mathcal{M}_9 &=\mathcal{O}(\alpha_{NB,NW})\,,\\
 %
\end{align}
Obviously, within our approximation this result agrees with Eq.~\eqref{eq:div2},
which only depends on $\alpha_{NB}$ and $\alpha_{NW}$ (and on $\alpha_{HNe}$ through $Y_e$).

%%%%%%%%%%%%%%%%%%%%%%%%%%%%%%
\subsection{$\overline{\nu_L}N\to B H_0$}
\begin{figure}[t]
 \centering
 \vspace{1cm}
  \includegraphics[height=3cm]{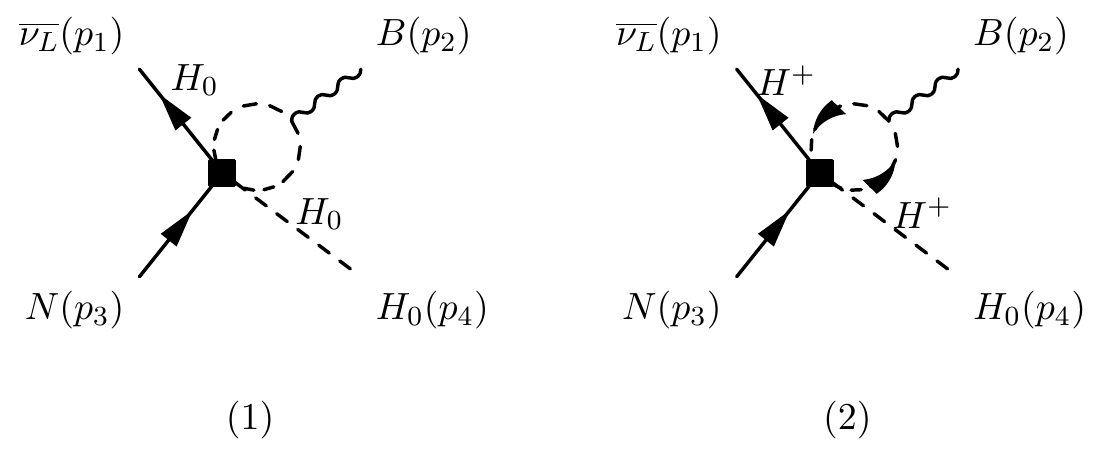}
 \caption{\it Relevant Feynman diagrams for $\overline{\nu_L}N\to B H_0$.}
 \label{fig:amp3}
\end{figure}

%%%%%%%%%%%%%%%%%%%%%%%%%%%%%%
%
The two relevant diagrams are those in Fig.~\ref{fig:amp3}. We obtain trivially
\begin{align}
 \mathcal{M}_1 &= 0\,,\\
 \mathcal{M}_2 &= 0\,.
 %
 %\mathcal{M}_3 &=\mathcal{O}(\alpha_{NB,NW})\,,\\
 %
 %\mathcal{M}_4 &=\mathcal{O}(\alpha_{NB,NW})\,,\\
 %
 %\mathcal{M}_5 &=\mathcal{O}(\alpha_{NB,NW})\,,\\
 %
 %\mathcal{M}_6 &= 0\,,\\
 %
 %\mathcal{M}_7 &=\mathcal{O}(\alpha_{NB,NW})\,,\\
 %
 %\mathcal{M}_8 &=\mathcal{O}(\alpha_{NB,NW})\,,\\
 %
 %\mathcal{M}_9 &=\mathcal{O}(\alpha_{NB,NW})\,,\\
 %
 %\mathcal{M}_{10} &=\mathcal{O}(\alpha_{NB,NW})\,,\\
 %
 %\mathcal{M}_{11} &=\mathcal{O}(\alpha_{NB,NW})\,,\\
 %
 %\mathcal{M}_{12} &=\mathcal{O}(\alpha_{NB,NW})\,,\\
 %
 %\mathcal{M}_{13} &=\mathcal{O}(\alpha_{NB,NW})\,,\\
 %
 %\mathcal{M}_{14} &=\mathcal{O}(\alpha_{NB,NW})\,,\\
 %
 %\mathcal{M}_{15} &=\mathcal{O}(\alpha_{NB,NW})\,.
 %
\end{align}
This again matches Eq.~\eqref{eq:div3} given our approximations.

%%%%%%%%%%%%%%%%%%%%%%%%%%%%%%
\subsection{$\overline{e_L}N\to W^3 H^+$}
\begin{figure}[t]
 \centering
  \includegraphics[height=3cm]{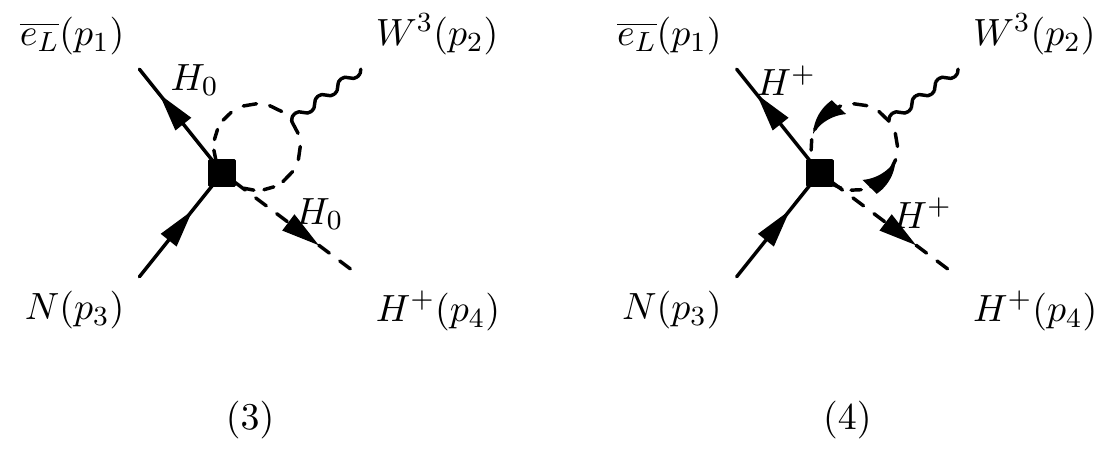}
 \caption{\it Relevant Feynman diagrams for $\overline{e_L}N\to W^3 H^+$.}
 \label{fig:amp4}
\end{figure}

%%%%%%%%%%%%%%%%%%%%%%%%%%%%%%
%
In this case we have the two diagrams of Fig.~\ref{fig:amp4}, which lead to
\begin{align}
 %
 %\MM_{1} &= \mathcal{O}(\alpha_{NB,NW})\,,\\
 %
 %\MM_{2} &= \mathcal{O}(\alpha_{NB,NW})\,,\\
 %
 \MM_{3} &= 0\,,\\
 \MM_{4} &= 0\,,
 %
 %\MM_{5} &= \mathcal{O}(\alpha_{NB,NW})\,,\\
 %
 %\MM_{6} &= \mathcal{O}(\alpha_{NB,NW})\,,\\
 %
 %\MM_{7} &= \mathcal{O}(\alpha_{NB,NW})\,,\\
 %
 %\MM_{8} &= \mathcal{O}(\alpha_{NB,NW})\,,\\
 %
 %\MM_{9} &= \mathcal{O}(\alpha_{NB,NW})\,,\\
 %
 %\MM_{10} &= \mathcal{O}(\alpha_{NB,NW})\,,\\
 %
 %\MM_{11} &= \mathcal{O}(\alpha_{NB,NW})\,,\\
 %
 %\MM_{12} &= \mathcal{O}(\alpha_{NB,NW})\,,\\
 %
 %\MM_{13} &= \mathcal{O}(\alpha_{NB,NW})\,,\\
 %
 %\MM_{14} &= \mathcal{O}(\alpha_{NB,NW})\,,\\
 %
 %\MM_{15} &= \mathcal{O}(\alpha_{NB,NW})\,,\\
 %
 %\MM_{16} &= \mathcal{O}(\alpha_{NB,NW})\,,\\
 %
 %\MM_{17} &= \mathcal{O}(\alpha_{NB,NW})\,,\\
 %
 %\MM_{18} &= \mathcal{O}(\alpha_{NB,NW})\,,\\
 %
 %\MM_{19} &= \mathcal{O}(\alpha_{NB,NW})\,.\\
 %
\end{align}
in agreement (within our approximation) with Eq.~\eqref{eq:div4}.

%%%%%%%%%%%%%%%%%%%%%%%%%%%%%%
\subsection{$\overline{N}N\to H_0^* H_0$}
%%%%%%%%%%%%%%%%%%%%%%%%%%%%%%
%
We have eleven relevant Feynman diagrams for this amplitude. They are depicted in Fig.~\ref{fig:amp5}.
The different contributions read:
\begin{figure}[t]
 \centering
  \includegraphics[height=3cm]{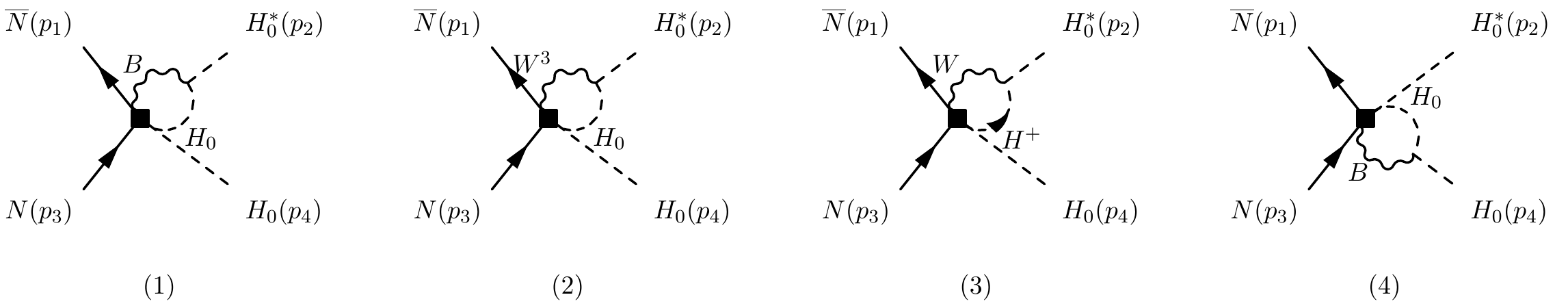}\\[0.4cm]
  \includegraphics[height=3cm]{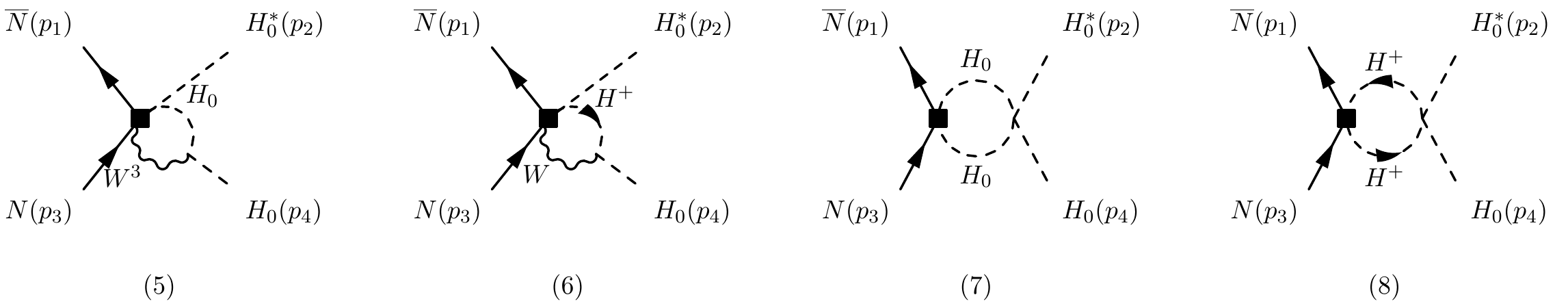}\\[0.4cm]
  \includegraphics[height=3cm]{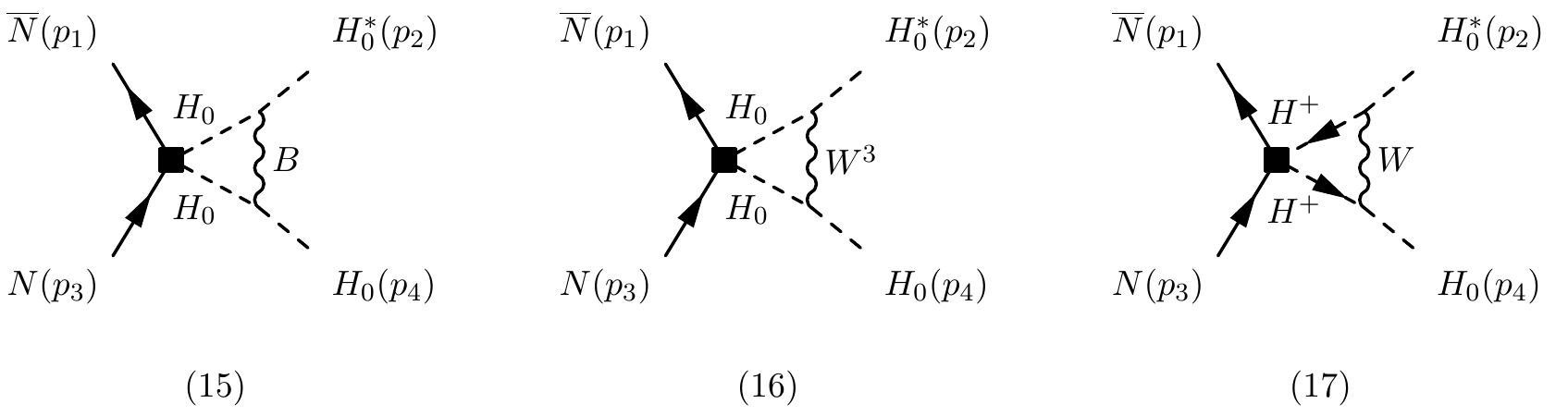}
 \caption{\it Relevant Feynman diagrams for $\overline{N}N\to H_0^* H_0$.}
 \label{fig:amp5}
\end{figure}
\begin{align}
 i\MM_{1} &= g_1^2 (A_2-B_3) \alpha_{HN} \overline{v_1}\slashed{p}_2 P_R u_3\,,\\
 i\MM_{2} &= g_2^2 (A_2-B_3)\alpha_{HN} \overline{v_1}\slashed{p}_2 P_R u_3\,,\\
 i\MM_{3} &= g_2^2(A_2+2 B_3)\alpha_{HN} \overline{v_1}\slashed{p}_2 P_R u_3\,,\\
 i\MM_{4} &= -\frac{g_1^2}{2}(A_2+2 B_3)\alpha_{HN} \overline{v_1}\slashed{p}_4 P_R u_3\,,\\
 i\MM_{5} &= -\frac{g_2^2}{2}(A_2+2 B_3)\alpha_{HN} \overline{v_1}\slashed{p}_4 P_R u_3\,,\\
 i\MM_{6} &= -g_2^2(A_2+2 B_3)\alpha_{HN} \overline{v_1}\slashed{p}_4 P_R u_3\,,\\
 i\MM_{7} &= 2\lambda_H (A_2-4 B_3)\alpha_{HN} \overline{v_1}(\slashed{p}_2 +\slashed{p}_4)P_R u_3\,,\\
 i\MM_{8} &= \lambda_H (A_2-4 B_3)\alpha_{HN} \overline{v_1}(\slashed{p}_2 +\slashed{p}_4)P_R u_3\,,\\
 %
 %\MM_{9} &= 0\,,\\
 %
 %\MM_{10} &= 0\,,\\
 %
 %\MM_{11} &= 0\,,\\
 %
 %\MM_{12} &= 0\,,\\
 %
 %\MM_{13} &= 0\,,\\
 %
 %\MM_{14} &= 0\,,\\
 %
 i\MM_{15} &= -\frac{g_1^2}{4}A_2\alpha_{HN} \overline{v_1}(\slashed{p}_2-\slashed{p}_4) P_R u_3\,,\\
 i\MM_{16} &= -\frac{g_2^2}{4}A_2\alpha_{HN} \overline{v_1}(\slashed{p}_2-\slashed{p}_4) P_R u_3\,,\\
 i\MM_{17} &= -\frac{g_2^2}{2}A_2\alpha_{HN} \overline{v_1}(\slashed{p}_2-\slashed{p}_4) P_R u_3\,.
\end{align}

Summing over all them we arrive at
\begin{align}
 i\mathcal{M}_\mathrm{loop} &= \frac{i}{32\pi^2\epsilon} 
 \left(g_1^2 + 3g_2^2\right) \alpha_{HN} 
 \overline{v_1} \left(\slashed{p}_2 - \slashed{p}_4\right) P_R u_3\,.
\end{align}
In this case, $p_1^\mathrm{FC} = p_1^\mathrm{QG}$,
$p_2^\mathrm{FC} = p_3^\mathrm{QG}$,
$p_3^\mathrm{FC} = -p_2^\mathrm{QG}$ 
and $p_4^\mathrm{FC} = -p_4^\mathrm{QG}$. 
Using this information, we match precisely Eq.~\eqref{eq:div5}.

%%%%%%%%%%%%%%%%%%%%%%%%%%%%%%
\subsection{$\overline{\nu_L}N H_0^*\to H_0^* H_0$}
%%%%%%%%%%%%%%%%%%%%%%%%%%%%%%
%
21 diagrams need to be computed in this case. They are all shown in Fig.~\ref{fig:amp6}. We have:
\begin{figure}[t!]
 \centering
  \hspace*{-1.5cm}
  \includegraphics[height=3cm]{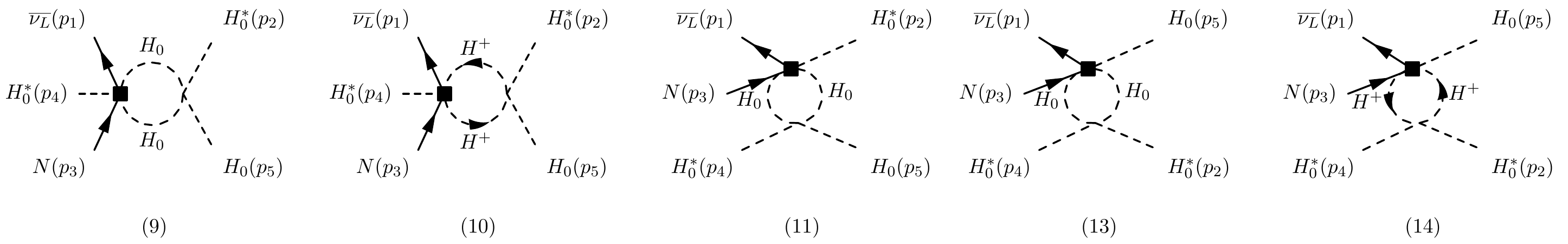}\\[0.4cm]
  \includegraphics[height=3cm]{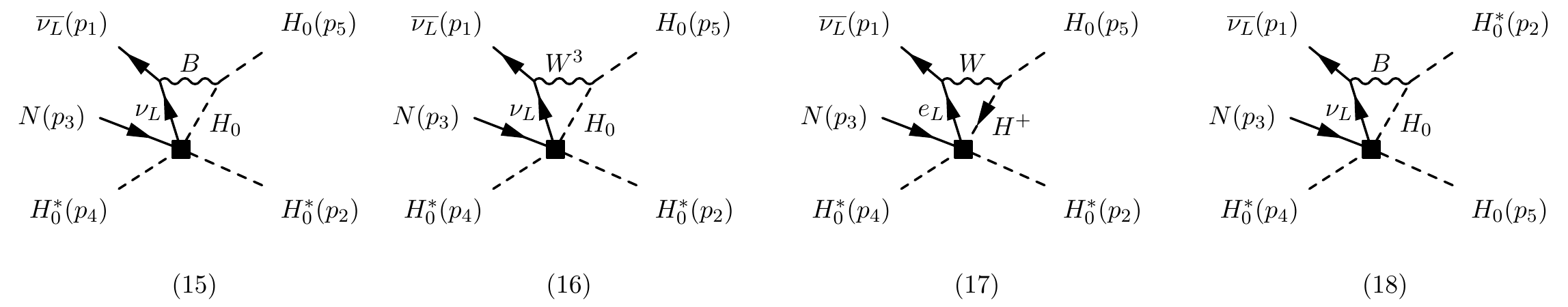}\\[0.4cm]
  \includegraphics[height=3cm]{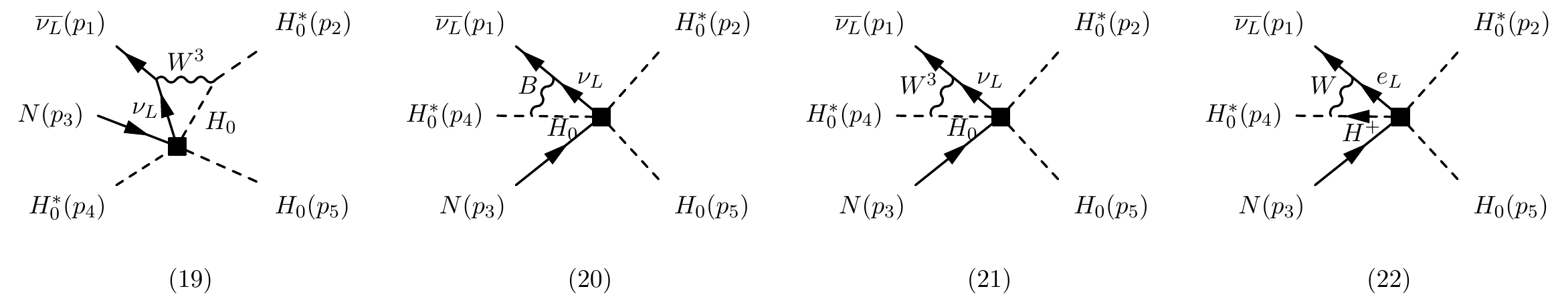}\\[0.4cm]
  \includegraphics[height=3cm]{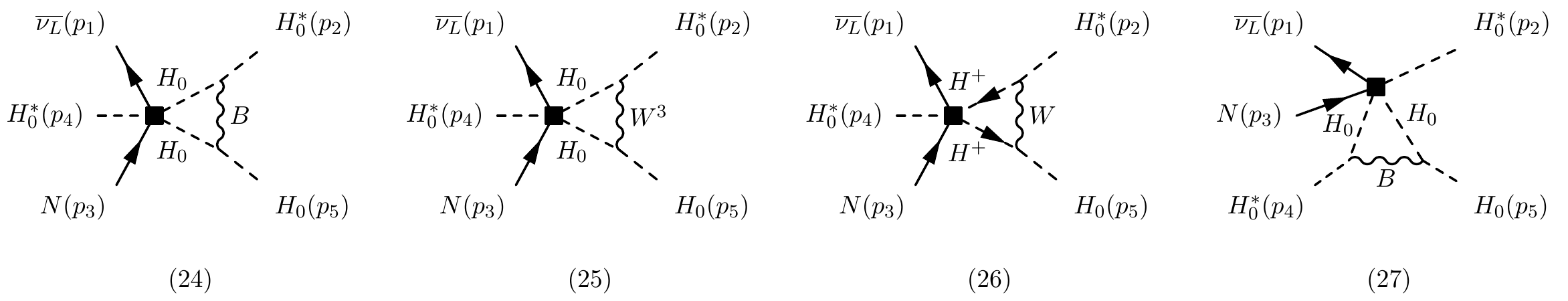}\\[0.4cm]
  \includegraphics[height=3cm]{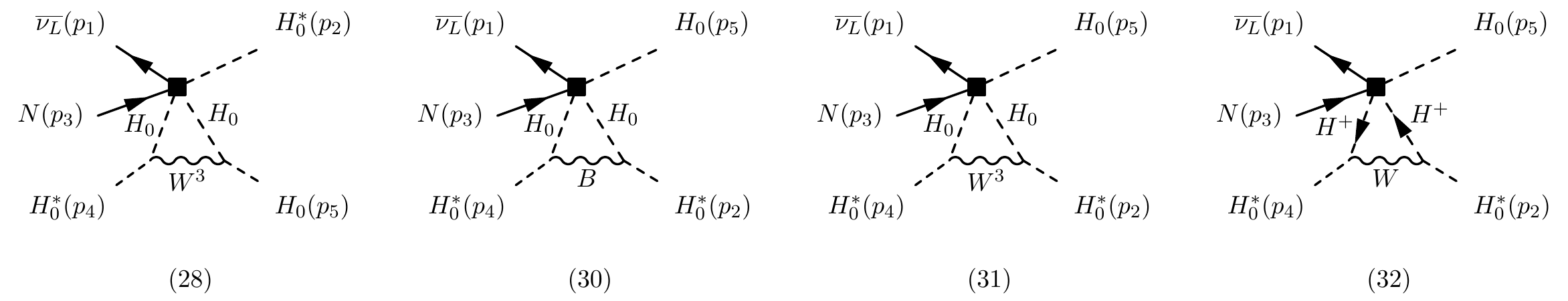}
 \caption{\it Relevant Feynman diagrams for $\overline{\nu_L} N H_0^*\to H_0^* H_0$.}
 \label{fig:amp6}
\end{figure}
\begin{align}
 %
 %\MM_1 &= 0\,,\\
 %
 %\MM_2 &= 0\,,\\
 %
 %\MM_3 &= 0\,,\\
 %
 %\MM_4 &= 0\,,\\
 %
 %\MM_5 &= 0\,,\\
 %
 %\MM_6 &= 0\,,\\
 %
 %\MM_7 &= 0\,,\\
 %
 %\MM_8 &= 0\,,\\
 %
 %\MM_9 &= 4\MM_{10},,\\
 i\MM_9 &= 4\lambda_H A_2 \alpha_{LNH}\overline{v_1} P_R u_3\,,\\
 %
 %i\MM_{10} &= \lambda_H A_2 \alpha_{LNH}\overline{v_1} P_R u_3\,,\\
 \MM_{10} &= \frac{1}{4}\MM_9\,,\\
 %
 %\MM_{11} &= 2\MM_{10}\,,\\
 \MM_{11} &= \frac{1}{2}\MM_9\,,\\
 %
 %\MM_{12} &= 0\,,\\
 %
 %\MM_{13} &= 4\MM_{10}\,,\\
 \MM_{13} &= \MM_{9}\,,\\
 %
 %\MM_{14} &= \MM_{10}\,,\\
 \MM_{14} &= \frac{1}{4}\MM_9\,,\\
 i\MM_{15} &= -\frac{g_1^2}{2}A_2\alpha_{LNH}\overline{v_1}P_R u_3\,,\\
 i\MM_{16} &= -\frac{g_2^2}{2}A_2\alpha_{LNH}\overline{v_1}P_R u_3\,,\\
 \MM_{17} &= \MM_{16}\,,\\
 \MM_{18} &= -\MM_{15}\,,\\
 \MM_{19} &= -\MM_{16}\,,\\
 \MM_{20} &= \MM_{15}\,,\\
 \MM_{21} &= \MM_{16}\,,\\
 \MM_{22} &= \MM_{16}\,,\\
 %
 %\MM_{23} &= \mathcal{O}(\alpha_{NB,NW})\,,\\
 %
 \MM_{24} &= \MM_{15}\,,\\
 \MM_{25} &= \MM_{16}\,,\\
 \MM_{26} &= \MM_{16}\,,\\
 \MM_{27} &= -\MM_{15}\,,\\
 \MM_{28} &= -\MM_{16}\,,\\
 %
 %\MM_{29} &= \mathcal{O}(\alpha_{NB,NW})\,,\\
 %
 \MM_{30} &= \MM_{15}\,,\\
 \MM_{31} &= \MM_{16}\,,\\
 \MM_{32} &= \MM_{16}\,.
\end{align}

As a result, we obtain
\begin{align}
 i\mathcal{M}_\mathrm{loop} &= \frac{i}{16\pi^2\epsilon} 
 \left(12\lambda_H - g_1^2 - 3g_2^2\right) \alpha_{LNH} 
 \overline{v_1} P_R u_3\,,
\end{align}
in agreement (within our approximation) with Eq.~\eqref{eq:div6}.

%%%%%%%%%%%%%%%%%%%%%%%%%%%%%%
\subsection{$\overline{N}e_R\to H^- H_0^*$}
\begin{figure}[t]
 \centering
  \includegraphics[height=3cm]{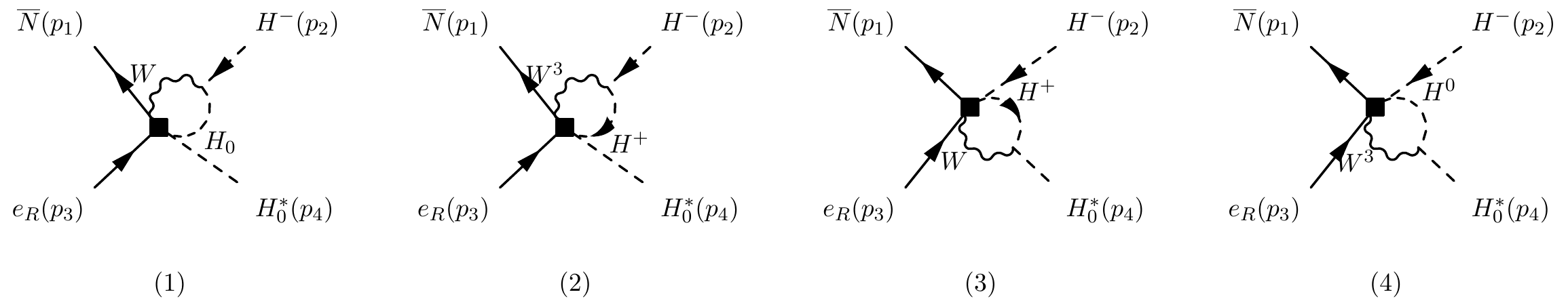}\\[0.4cm]
  \includegraphics[height=3cm]{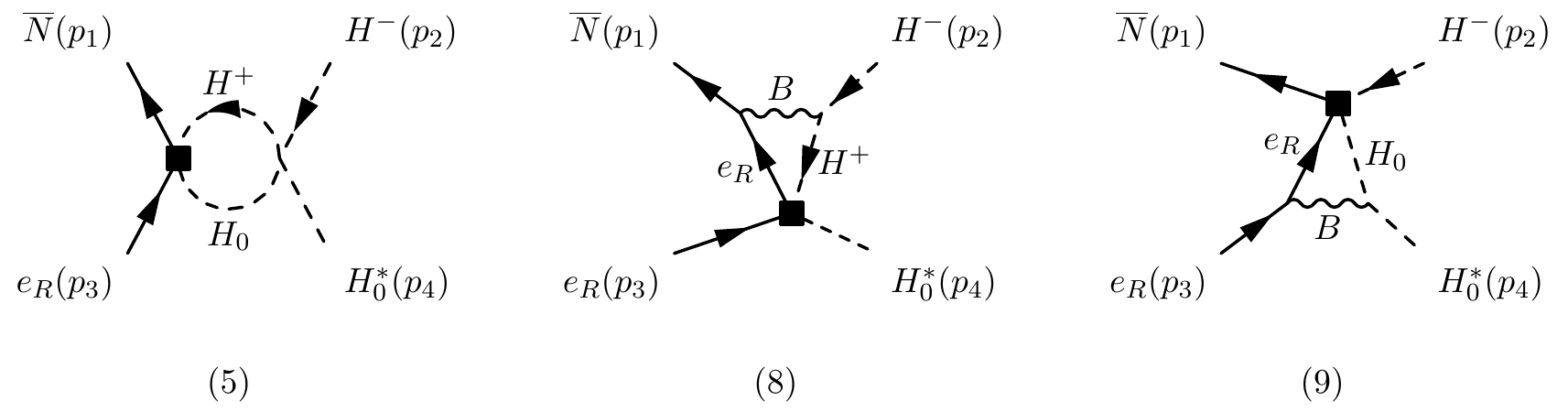}\\[0.4cm]
  \includegraphics[height=3cm]{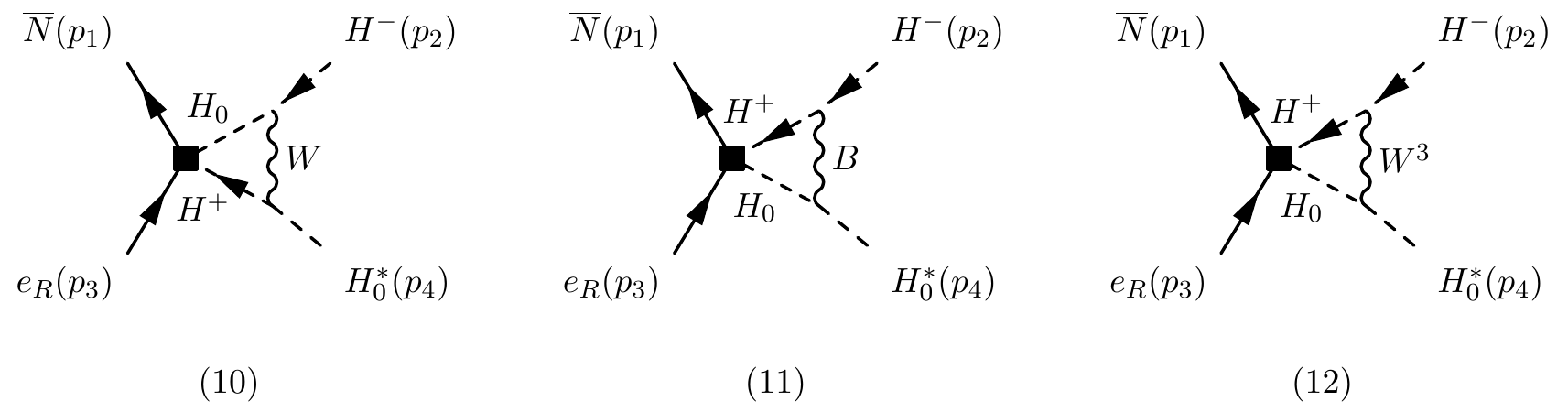}
 \caption{\it Relevant Feynman diagrams for $\overline{N}e_R\to H^- H_0^*$.}
 \label{fig:amp7}
\end{figure}

%%%%%%%%%%%%%%%%%%%%%%%%%%%%%%
%
Finally, this amplitude splits into the ten diagrams of Fig.~\ref{fig:amp7}. We have

\begin{align}
 i\mathcal{M}_1 &= 2 g_2^2 \left(A_2 - B_3\right) \alpha_{HNe}
 \overline{v_1} \slashed{p}_2 P_R u_3\,, \\
 \mathcal{M}_2 & = \frac{1}{2}\mathcal{M}_1\,, \\
 i\mathcal{M}_3 &= -2 g_2^2 \left(A_2 - B_3\right) \alpha_{HNe}
 \overline{v_1} \slashed{p}_4 P_R u_3\,, \\
 \mathcal{M}_4 &= \frac{1}{2}\mathcal{M}_3\,, \\
 i\mathcal{M}_5 &= -\lambda_H \left(A_2 - 4B_3\right) \alpha_{HNe} 
 \overline{v_1} \left(\slashed{p}_2 + \slashed{p}_4\right) P_R u_3\,, \\
 %
 %\mathcal{M}_6 &= \mathcal{O}\left(\frac{\alpha_{HN}\alpha_{HNe}}{\Lambda^4}\right), \\
 %
 %\mathcal{M}_7 &= \mathcal{O}\left(\frac{\alpha_{HN}\alpha_{HNe}}{\Lambda^4}\right), \\
 %
 i\mathcal{M}_8 &= - g_1^2  \alpha_{HNe} 
 \overline{v_1} \left[\left(A_2+B_3\right)\slashed{p}_2 
 + \frac{1}{2} A_2 \left(\slashed{p}_3-\slashed{p}_4\right)\right] P_R u_3\,, \\
 i\mathcal{M}_9 &= g_1^2 \alpha_{HNe} 
 \overline{v_1} \left[\left(A_2+B_3\right)\slashed{p}_4 
 + \frac{1}{2} A_2 \left(\slashed{p}_3-\slashed{p}_2\right)\right] P_R u_3\,, \\
 i\mathcal{M}_{10} &= -\frac{1}{2} g_2^2  A_2 \alpha_{HNe}
 \overline{v_1} \left(\slashed{p}_2-\slashed{p}_4\right) P_R u_3\,, \\
 i\mathcal{M}_{11} &= \frac{1}{4} g_1^2  A_2 \alpha_{HNe} 
 \overline{v_1} \left(\slashed{p}_2-\slashed{p}_4\right) P_R u_3\,, \\
 \mathcal{M}_{12} &%= -\frac{1}{4} \alpha_{HNe} g_2^2  A_2
 %\overline{v_1} \left(\slashed{p}_2-\slashed{p}_4\right) P_R u_3 
 = \frac{1}{2}  \mathcal{M}_{10}\,.
\end{align}

Thus, we finally obtain:
\begin{align}
 i\mathcal{M}_\mathrm{loop} &= -\frac{3i}{32\pi^2\epsilon} 
 \left(g_1^2 - g_2^2\right) \alpha_{HNe} 
 \overline{v_1} \left(\slashed{p}_2-\slashed{p}_4\right) P_R u_3\,.
\end{align}
In this case, $p_1^\mathrm{FC} = p_1^\mathrm{QG}$,
$p_2^\mathrm{FC} = p_3^\mathrm{QG}$,
$p_3^\mathrm{FC} = -p_2^\mathrm{QG}$, 
and $p_4^\mathrm{FC} = -p_4^\mathrm{QG}$. So we easily see that we match Eq.~\eqref{eq:div7}.

%%%%%%%%%%%%%%%%%%%%%%%%%%%%%%
%\bibliographystyle{JHEP}
\bibliography{NSMEFT_RGEs_v2}
%%%%%%%%%%%%%%%%%%%%%%%%%%%%%%

%%%%%%%%%%%%%%%%%%%%%%%%%%%%%%
\end{document}